\documentclass[aps,prb,superscriptaddress,showpacs,citeautoscript,twocolumn]{revtex4-1}

\usepackage{graphicx}
\usepackage{amsmath}
\usepackage{amssymb}
\usepackage{color}
\newcommand{\bea}{\begin{eqnarray*}}
\newcommand{\eea}{\end{eqnarray*}}
\newcommand{\bne}{\begin{equation*}}
\newcommand{\ede}{\end{equation*}}

\newcommand{\bnen}{\begin{equation}}
\newcommand{\eden}{\end{equation}}
\newcommand{\bean}{\begin{eqnarray}}
\newcommand{\eean}{\end{eqnarray}}
\newcommand{\bnsn}{\begin{subequations}}
\newcommand{\edsn}{\end{subequations}}

\newcommand{\bna}{\begin{array}}
\newcommand{\eda}{\end{array}}
\newcommand{\bnm}{\begin{enumerate}}
\newcommand{\edm}{\end{enumerate}}

\renewcommand{\vec}[1]{\text{\boldmath{$ #1 $}}}

\begin{document}

\title{Electrical tuning of Rashba spin-orbit interaction in multigated InAs nanowires}

\author{Zolt\'an Scher\"ubl}
%\email{scherubl.zoltan@gmail.com}
\affiliation{Department of Physics, Budapest University of Technology and Economics and
Condensed Matter Research Group of the Hungarian Academy of Sciences, 1111 Budapest, Budafoki \'ut 8., Hungary}
\author{Gerg\H{o} F\"ul\"op}
\affiliation{Department of Physics, Budapest University of Technology and Economics and
Condensed Matter Research Group of the Hungarian Academy of Sciences, 1111 Budapest, Budafoki \'ut 8., Hungary}
\author{Morten H. Madsen}
\affiliation{Center for Quantum Devices, Niels Bohr Institute, University of Copenhagen, 2100 Copenhagen, Denmark}
\author{Jesper Nyg\r{a}rd}
\affiliation{Center for Quantum Devices, Niels Bohr Institute, University of Copenhagen, 2100 Copenhagen, Denmark}
\author{Szabolcs Csonka}
%\email{csonka@dept.phy.bme.hu}
\affiliation{Department of Physics, Budapest University of Technology and Economics and
Condensed Matter Research Group of the Hungarian Academy of Sciences, 1111 Budapest, Budafoki \'ut 8., Hungary}

\date{\today}

\begin{abstract}
Indium arsenide (InAs) nanowires (NWs) are a promising platform to fabricate quantum electronic devices, among others they have strong spin-orbit interaction (SOI). The controlled tuning of the SOI is desired in spin based quantum devices. In this study we investigate the possibility of tuning the SOI by electrostatic field, which is generated by a back gate and two side gates placed on the opposite sides of the NW. The strength of the SOI is analyzed by weak anti-localization effect. We demonstrate that the strength of SOI can be strongly tuned by a factor of 2 with the electric field across the NW, while the average electron density is kept constant. Furthermore a simple electrostatic model is introduced to calculate the expected change of SOI. Good agreement is found between the experimental results and the estimated Rashba type SOI generated by the gate-induced electric field.
\end{abstract}

 %Max 4
%71.70.Ej	Level splitting and interactions / Spin-orbit coupling, Zeeman and Stark splitting, Jahn-Teller effect
%72.25.Rb Spin polarized transport / Spin relaxation and scattering
%73.20.Fz Electron states at surfaces and interfaces / Weak or Anderson localization
%73.63.-b	Electronic transport in nanoscale materials and structures
%73.63.Nm Quantum wires
%73.63.Fg Nanotubes
%85.35.-p Nanoelectronic devices

\maketitle

\section{Introduction}

Recently III-V semiconductor nanowires (NWs) have attracted increasing attention in the field of quantum electronics. The strong spin-orbit interaction (SOI) is an attractive property of InAs and InSb NWs:  it results large g-factor, which leads to large Zeeman-splitting \cite{JespersenPRB2006,SchroerPRL2011,CsonkaNanoLett2008,dHollosyAIPConfProc2013}, it allows to effectively address spins by electric fields (EDSR) \cite{PeterssonNature2012,StehlikPRL2014}, furthermore to define new types of quantum bits, so called spin-orbit qubits \cite{NadjPergeNature2010,NadjPergePRL2012,BergPRL2013}, where the quantum information is stored by coupled spin and orbital degree of freedom. Very recently theoretical proposals suggested \cite{AliceaRepProgPhys2012,BeenakkerAnnRevCondMatPhys2013,FuPRL2008,SauPRl2010,AliceaPRB2010,LutchynPRL2010,OregPRL2010} that these one dimensional NWs are an ideal platform to induce topological superconductivity, and they may host pairs of so-called Majorana bound states, which are key building blocks for topological quantum computation. The strong SOI is also a key ingredient of these Majorana wires. The models assume a Rashba-type SOI, which in combination with magnetic field leads to the essential locking of spin and momentum in the 1D band structure. Several groups have reported the signature of these elusive particles \cite{MourikScience2012,DasNatPhys2012,DengNanoLett2012,FinckPRL2013,ChurchillPRB2013} as a presence of a zero bias anomaly. Since the experimental conditions needed to reach this topological superconducting state are challenging, it would be highly desirable to have further experimental control parameters to analyze and tune the Majorana wires, e.g. a tunable Rashba SOI strength could give a better insight.

In this paper we demonstrate the tunability of the Rashba SOI strength by means of local gating. We have fabricated InAs NW devices with pairs of local side gates (SGs)  (see Fig. 1a\&b). By applying voltages with opposite sign on these two SGs an electrostatic field can be generated. The influence of the electrostatic field on the strength of SOI is studied via weak anti-localization (WAL) measurements.

WAL arises from interference effects \cite{AltshulerJETPLett1981,ChakravartyPhysRep1986,BeenakkerPRB1988,KurdakPRB1992,BergmannPhysRep1984,BergmannSSC1982,ChakravartyPhysRep1986}, where the time-reversed electron path pairs interfere destructively in zero magnetic field leading to enhanced conductance. Breaking the time-reversal symmetry of the system, i.e. by applying a magnetic field, suppresses the interference term of the conductance, resulting in negative magnetoconductance (MC).
WAL was used  to extract the strength of SOI in InAs and InSb NWs \cite{HansenPRB2005,DharaPRB2009,LiangNanoLett2009,RoulleauPRB2010,HernandezPRB2010,LiangPRB2010,LiangNanoLett2012, WeperenPRB2015}.
In most of the studies the WAL signal was investigated in field effect transistor (FET) configuration, where a global back gate (BG) electrode is placed under a silicon oxide layer, which changes the electron density and the electrostatic field in the NW at the same time \cite{HansenPRB2005,DharaPRB2009,LiangNanoLett2009,RoulleauPRB2010,LiangPRB2010,HernandezPRB2010}. Other studies introduces a top gate, beside the BG, separated from the NW by a solid electrolyte, PMMA \cite{LiangNanoLett2012} or a HfO$_2$ layer \cite{WeperenPRB2015}.

Here we report results on InAs NWs using a new geometry where two SGs are added beside the BG. In this way similar to FET structure the BG serves to change the electron density, while the SGs could be used to induce a transverse electric field in the wire by applying opposite voltages without changing the overall electron density.

Here we show that the total SOI strength can be enhanced with a factor of 2, only by applying electric field in the NW. Furthermore a simple electrostatic model is introduced which shows that this tuning is consistent with the external field-induced Rashba-type SOI.

\section{Device and Methods}
\label{sec:devmet}

 InAs NWs were grown by molecular beam epitaxy \cite{AagesenNatNanotech2007}. After growth, they were deposited onto a thermally oxidized, degenerately doped n-Si substrate from a 2-propanol dispersion. The thickness of the SiO$_2$ layer is 400~nm and the underlying doped Si layer serves as the global BG electrode. The source (S), drain (D) and SG (SG1\&2) electrodes were defined in a two step e-beam lithography process and were deposited by UHV e-beam evaporation (Ti/Au 10/90~nm), after Ar ion beam etching to remove the native oxide from the surface of the NW. A false color SEM picture of the sample is shown on Fig.~1a. The SG separation is 220~nm, the separation of the S and D electrodes is $L=1$~$\mu$m and the diameter of the NW is $W=77$~nm. Fig.~1b shows the cross section of the device perpendicular to the NW at the center of the SG electrodes on Fig.~1a.

In contrast to previously used top gated geometries, an important advantage of the side gated sample is that the gate electrode is not in direct contact with the NW. This way formation of charge traps at the interface of the NW and the top gate electrode is prevented. Further advantage is that electric field can be induced in the NW without significant change of the conductance by applying opposite voltages on the SGs.

Low temperature transport measurements were performed in a liquid helium cryostat with variable temperature insert system which is equipped with a superconducting magnet. Prior to cool down the sample was pumped overnight at room temperature to remove the water contamination from the surface of the NW.

The conductance, $G$ follows a typical n-type FET characteristic of the NW, as it is shown on Fig.~1c at $T=50$~K. Both SGs and the BG can individually increase the conductance up to 1.5~$G_0$ ($G_0=2e^2/h$ is the conductance quantum), with a saturation like tendency, and can completely deplete the NW, i.e. quench the conductance, while the other gates are kept at zero potential. The transconductance curves measured as a function of $V_{SG1}$ and $V_{SG2}$ (black and red curves on Fig.~1c) are practically overlapping, which indicates that the capacitive coupling to the SGs are almost equal. The SGs have a 4 times stronger effect than the BG due to closer spacing to the NW (70~nm, 400~nm respectively). Note that different horizontal axes are used for the BG and the SGs.

To determine the strength of spin relaxation the magnetoconductance of the device was measured in various gate settings at $T=4.2$~K. The magnetic field was perpendicular to the substrate (see Fig.~1b). In the used gate geometry with 70~nm spacing between the NW and SGs, SG voltages ($V_{SGi}$) up to 20~V and difference of 30~V between $V_{SG1}$ and $V_{SG2}$ could be applied without electric breakdown.

	\begin{figure}[!htbp]
	\begin{center}
	\includegraphics[width=8.5cm]{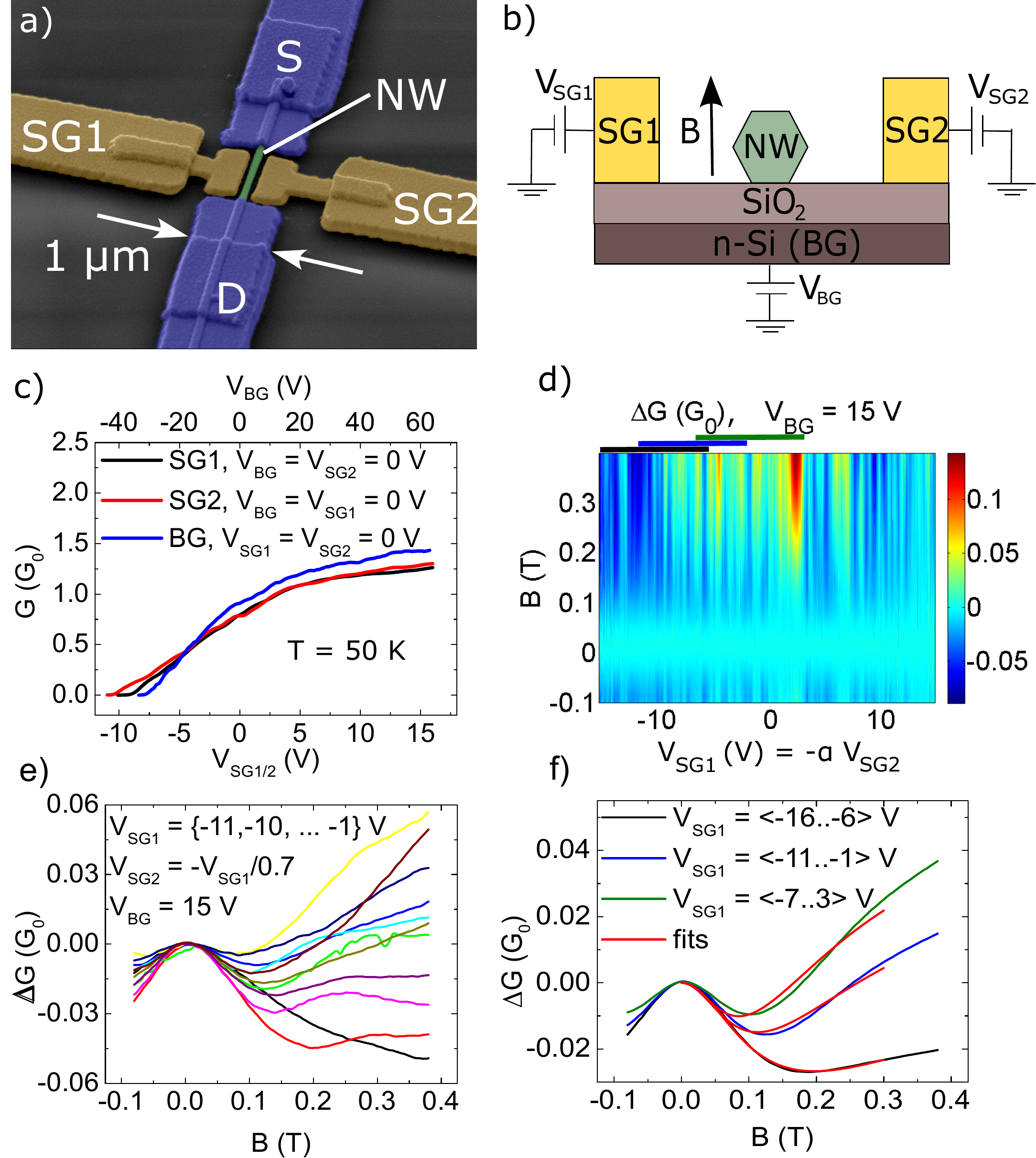}
	\caption{a) False color SEM picture of a representative device.
	b) Cross section of the device, perpendicular to the nanowire (NW) axis  along the center of side gate electrodes. Two side gates (SG1\&2) and a back gate (BG) electrode can be used to apply electric field on the NW. Beside that magnetic field perpendicular to the substrate and NW was used.
	c) The conductance of the NW as a function of different gate voltages at T = 50~K. Note that the \textit{x} axis for BG and SG curves are different.
	d) Variation of magnetoconductance (MC) i.e. $\Delta G(B)=G(B)-G(B=0)$ as a function of the asymmetric SG voltage. For the value of $\alpha$ see  Eq.~\ref{eq:vsg2}. 
	e) Typical MC curves, vertical cuts from panel $d)$. All curves show a peak around zero magnetic field due to the weak anti-localization (WAL) and oscillations at higher B due to universal conductance fluctuation (UCF).
	f) Averaged MC curves in a 10~V-wide gate voltage window are shown. 
Here $<.>$ indicates the gate range of averaging. The effect of SG-induced electric field can be seen. As stronger electric field is applied, the minimum of the WAL curve shifts to higher B field values and decreases, which indicates shorter $l_{SO}$, stronger SOI. Red curves are fits with theory of WAL (see Eq.~1). 
}
	\label{fig1}
	\end{center}
	\end{figure}

Changes of the MC in low magnetic field contains the spin relaxation-related WAL signal, therefore we focus on the variation of the conductance from the zero magnetic field value, i.e. $\Delta G=G(B)- G(B=0$~T$)$. Fig.~1d\&e show a typical set of measurements of $\Delta G$ as a function of $B$ and gate voltages.
The MC curves show two general features, first every curve has a local maximum around zero magnetic field, which is a signature of WAL. Second at higher magnetic fields ($B>0.1$~T$)$,  further oscillations appear in $\Delta G$ , which depend on the gate voltages in a random fashion. This random fluctuation (called universal conductance fluctuation, UCF) is due to the variation of the interference condition of coherent electron paths as $B$ or the gate voltage is changed.

The WAL signal allows us to extract important transport parameters, like the phase coherence length and the spin relaxation length \cite{AltshulerJETPLett1981,BeenakkerPRB1988}. To relieve the WAL signal from UCF the MC curves were averaged in moving gate voltage window \cite{RoulleauPRB2010} with a width of 10~V. Three averaged MC curves are shown in Fig.~1f, the corresponding gate voltage windows are indicated with colored bars in Fig.~1d, above the upper horizontal axis. Due to the averaging the strongly varying features at high magnetic fields disappear and only a characteristic peak around $B=0$~T remains with a dip and a monotonic increase towards higher $B$ field. This line shape is a characteristic WAL signal.

The averaged MC curves are fitted with the theoretical formula of WAL in one dimensional systems \cite{AltshulerJETPLett1981,BeenakkerPRB1988,KurdakPRB1992,LiangPRB2010},

\bean \label{eq:WAL} \Delta G(B) = &-& \frac{2e^2}{hL}\left[ \frac{3}{2} \left( \frac{1}{l^2_{\phi}}+\frac{4}{3l^2_{SO}}+\frac{1}{l^2_{B}} \right)^{-1/2} \right. \nonumber \\ &-& \left. \frac{1}{2} \left( \frac{1}{l^2_{\phi}}+\frac{1}{l^2_{B}} \right)^{-1/2} \right], \eean
where $l_{\phi}$ is the phase coherence length, $l_{SO}$ is the spin relaxation length, $L = 1$~$\mu$m is  the separation of S and D electrodes, $h$ is the Planck constant, $e$ is the electron-charge and
\bne l^2_{B} = \frac{8\hbar^2}{e^2B^2W^2\sqrt{3}}, \ede
where $B$ is the magnetic field and $W$ is the diameter of the NW. Note that the geometrical factor (i.e. the area of the cross section), $W^2$ in Ref.~\onlinecite{AltshulerJETPLett1981,BeenakkerPRB1988} has been replaced by $\frac{3\sqrt{3}}{8}W^2$ to adapt the formula to the hexagonal cross section of the NW, following the idea of Ref.~\onlinecite{LiangPRB2010}.

This formula is valid in the dirty limit, i.e. the elastic scattering length, $l_e$ is much smaller than the wire diameter, and in the low magnetic field limit, i.e. $l_m = \sqrt{\hbar/eB} >> W$. Our sample is close to fulfill the first condition, since the diameter of the NW is $W=77$~nm, and the elastic scattering length is $l_e \approx 10-20$~nm, determined from transconductance measurements. \cite{DuNanoLett2009,LiangPRB2010} The second inequality implies a $|B|<0.1$~T condition for the fitting. The MC curves were fitted in the $0<B<0.3$ T interval, since it is needed to contain the minimum of the WAL curve for a reliable fitting. The formula has two fitting parameters, the phase coherence length ($l_{\phi}$) and the spin relaxation length ($l_{SO}$). Assuming that the SOI dominates the spin relaxation, $l_{SO}$ is used to measure the strength of SOI\cite{ChakravartyPhysRep1986,HikamiPTP1980}. According to Eq.~1 a signature of enhanced SOI (i.e. reduced $l_{SO}$) is the shift of the minimum of the WAL curve to higher B values.

The three experimental curves presented in Fig.~1f show a clear shift of the minimum of the WAL signal as the asymmetric SG voltage is increased. It is in agreement with the expected enhancement of SOI as higher electric field is induced by larger gate voltage. In order to extract $l_{\phi}$ and $l_{SO}$ these curves were fitted by Eq.~1.
The fitted MC curves are shown with red lines in Fig.~1f. There is a reasonable agreement between the measured and fitted curves, even though the magnetic field window used for the fit is somewhat wider than the validity of the model. The extracted $l_{SO}$ parameters for the three plotted measurements are 175, 144 and 92~nm (from top to bottom), showing an enhancement of SOI in increasing electrostatic fields. In the following, using the fitting procedure described above, $l_\phi$ and $l_{SO}$ are extracted  from averaged MC traces measured at different gate settings. We present results measured on one particular device, but similar effects were observed on several other devices.

\section{Results}

\begin{figure}[!htbp]
	\begin{center}
	\includegraphics[width=8.5cm]{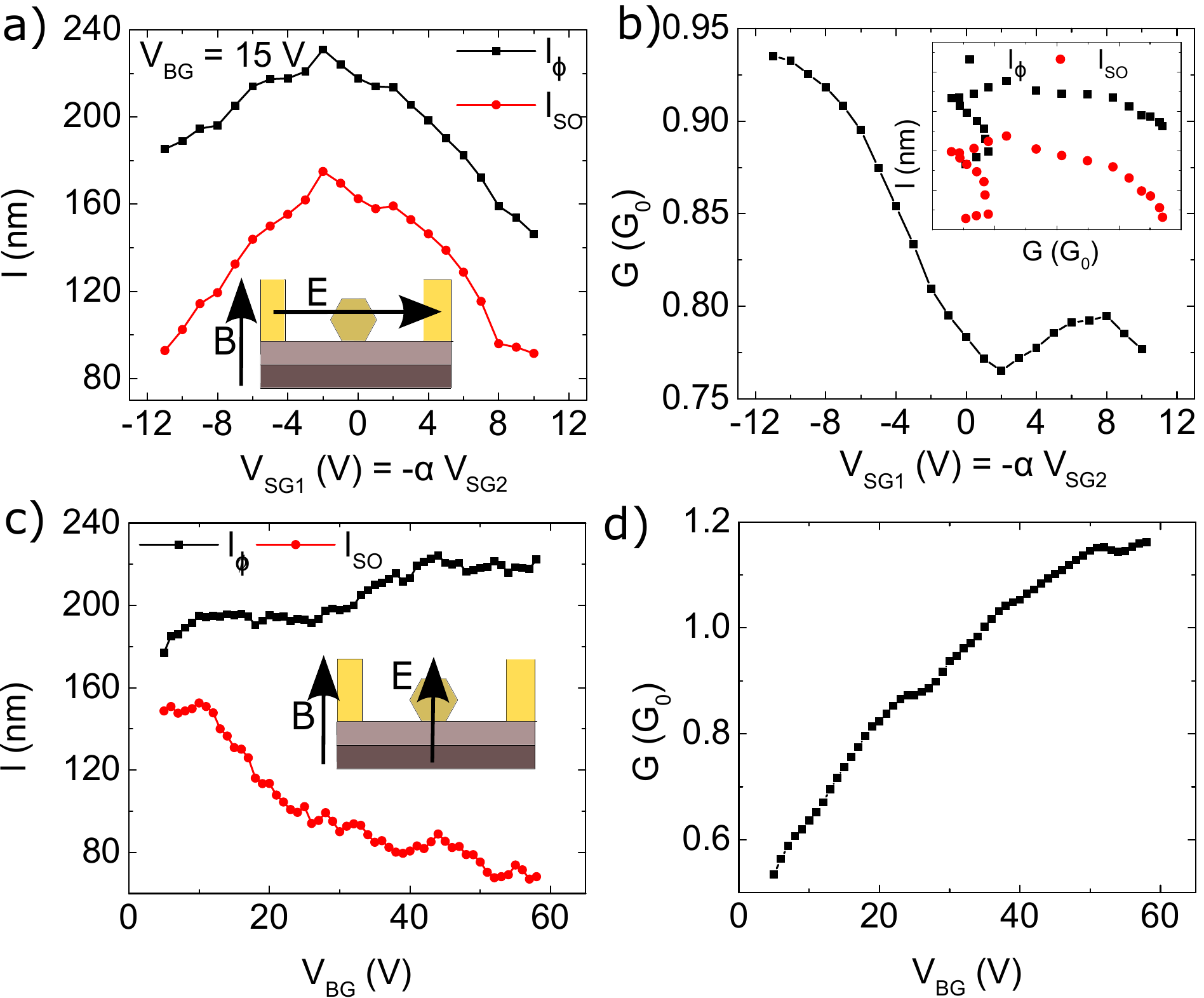}
	\caption{a) The fitted $l_{SO}$ and $l_{\phi}$ values as a function of the asymmetric side gate (SG) voltage. For the value of $\alpha$ see Eq.~\ref{eq:vsg2}. The peak-like reduction of $l_{SO}$ on the red curves indicates the tuning of SOI with a factor of 2, which is due to the electric field across the nanowire (NW). (inset) The geometry of the measurement with the direction of the electric and magnetic fields.
	b) The conductance of the NW at $B = 0$ changes less than 20\% while the asymmetric SG voltage is tuned. (inset) Combined data of a) and b), $l_{SO}$ and $l_{\phi}$ values as a function of the NW conductance. The lack of correlation of the conductance and the fitting parameters indicates that the tuning is not the result of the change of conductance.
	c),d) Same as a)\&b) for the back gate (BG) tuned measurement. At high positive BG voltage the electric field results in the reduction of $l_{SO}$ by a factor of 2, while the conductance and so the electron density is strongly increased.}
	\label{fig2}
	\end{center}
	\end{figure}

In our geometry (see Fig.~1a\&b) there are 3 gate electrodes which can induce an electric field inside the NW. In the following the influence of two different electric field profiles on $l_{SO}$ is studied. In the first case asymmetric potential is applied on the SGs, while the BG voltage is fixed. In the second case finite voltage is applied only on BG, while the SGs are grounded. In the following we show that the induced external electric field can strongly enhance the SOI in both cases.

\subsection{Tuning with in-plane electric field}
\label{sec:sg}

In the standard FET geometry the BG voltage can be used to induce electric field in the NW, however $V_{BG}$ changes the strength of the electric field and the electron density at the same time (see Fig. 1c). The main advantage of the 3-gate-configuration is that it allows to generate an electrostatic field, \textit{without} changing the conductance and the average electron density in the NW, by applying opposite voltages on the two SGs (see Fig. 2a inset). This asymmetric gating induces an electric field parallel to the substrate plane, perpendicular to the NW axis.

In order to keep the conductance of the NW constant while the electrostatic field was enhanced, the asymmetric SG voltages were swept in the following way: $V_{SG1} = - \alpha V_{SG2}$, where
\bnen \label{eq:vsg2} \alpha = \begin{cases} 0.7 & \text{if } V_{SG2}>0 \text{V} \\ 1/0.7 & \text{if } V_{SG2}<0 \text{V} \end{cases}, \eden
i.e. on the negatively charged SG electrode a voltage smaller by a factor of 0.7 was applied because of the nonlinear gate dependance of the conductance (see Fig.~1c).

In the measurement, shown on Fig.~2a\&b $V_{SG1}$ was swept between -16 and 15~V, but on the voltage axis the data is plotted in the [-11~V, 10~V] interval, since it was averaged in a window of 10~V. The BG voltage was fixed at 15 V to maintain the conductance above 0.5~$G_0$, which ensures the visibility of the WAL signal. Fig.~2b shows the $B = 0$~T conductance extracted from averaged MC measurements as a function of asymmetric SG voltages applied according to Eq.~2. The conductance was kept close to constant within 0.2~$G_0$.

On the main panel of Fig.~2a the $l_{\phi}$ and $l_{SO}$, determined by fitting the averaged MC curves with Eq.~\ref{eq:WAL}, are shown. At zero SG voltages the spin relaxation length is about 170~nm, which is comparable with earlier results \cite{HansenPRB2005,DharaPRB2009,RoulleauPRB2010,LiangNanoLett2012}. By applying the asymmetric voltage on the SGs, and thus generating an electric field perpendicular to the NW, $l_{SO}$ can be significantly reduced to ~90~nm. The MC data shown in Fig.~1d,e\&f also corresponds to this measurement.

The conductance also varied slightly  with the applied SG voltages. However plotting $l_{\phi}$ and $l_{SO}$ as a function of the conductance (Fig.~2b inset) reveals that there is no correlation between the conductance and the fitting parameters. It indicates that the reduction of $l_{SO}$ is not the result of the change of the conductance.

Increasing the electric field across the NW, the electron density is dominantly pushed to the surface of the NW (see Sec.\ref{sec:sim}), which can explain the reduction of the phase coherence length with increasing electric field (see the black curve in Fig.~2a), due to the increased scattering on the surface-roughness or residual contamination.

Investigating other devices the highest reduction factor of $l_{SO}$ was 3 for $V_{SG1} =18$~V.

\subsection{Tuning with perpendicular electric field}
\label{sec:bg}

In the previous section it was shown that $l_{SO}$ can be strongly tuned by an asymmetric SG induced in-plane electric field. In order to have a comparison, in a second set of measurements the BG was used as the source of the electric field, while the SG electrodes were grounded. This setting is similar to the standard FET-like gating \cite{HansenPRB2005,DharaPRB2009,LiangNanoLett2009,RoulleauPRB2010,HernandezPRB2010,LiangPRB2010}. The main difference compared to case of asymmetric SG voltages is that in addition to the strength of the electrostatic field, the electron density also changes with $V_{BG}$, resulting in the change of the conductance, as it is shown in Fig.~2d. Note that the conductance here is extracted from the averaged MC data. Also note that the transconductance curve is shifted to higher $V_{BG}$ values compared to the blue curve on Fig.~1c, since the high BG voltages (up to 60~V), required to achieve high electrostatic fields across the NW, caused the filling of charge traps in the SiO$_2$ in the close proximity of the NW, which screens the effect of the BG. The rearrangement of these static charges results a in difference between the extracted physical quantities at the same gate configuration for the two different set of measurements.

The corresponding fitted $l_{\phi}$ and $l_{SO}$ parameters are plotted in Fig.~2c. $l_{SO}$ shows a clear tendency as a function of $V_{BG}$. At small gate voltage values $l_{SO}$ is around 150~nm, while for larger $V_{BG}$ and thereby enhanced electrostatic field $l_{SO}$ monotonously decreases. Applying BG voltage of more than 60~V, $l_{SO}$ is reduced even to 70~nm.

In some of the previous studies investigating the SOI via WAL in standard FET geometry in InAs and InSb NWs no tuning was observed \cite{RoulleauPRB2010}, while in others a tuning effect up to 30\% was reported \cite{HansenPRB2005,DharaPRB2009}. All of these studies used smaller gate voltage range than us. A strong tuning, comparable with our results was only observed in devices using a top gate \cite{WeperenPRB2015} or electrolyte gate \cite{LiangNanoLett2012}. Our measured $l_{SO}$ values at low gate voltage values are in good agreement with these studies.

The change of the phase coherence length is not significant, in the used range of BG voltage $l_\phi$ increases less than 20\% from 180~nm to 220~nm. Interestingly its tendency is just opposite to the case of asymmetric SGs (see Fig. 2a). The gate voltage-induced enhancement can be explained with the reduction of electron-electron interaction at higher carrier densities (higher conductance).

\section{Simulation}
\label{sec:sim}

In the previous section it was shown that by applying an electric field on the NW with gate electrodes the SOI can be enhanced by a factor of 2. In this section a simple electrostatic model is introduced, which is used to calculate the electrostatic field ($\vec E(\vec r)$) profile inside the NW, and to estimate the electric field-induced Rashba SOI. A good agreement was found between the numerical results for $l_{SO}$ and the measured values presented in the previous section.

For simplification the measured device is modeled with its 2 dimensional cross section, shown in Fig.~1b and discussed in Sec.~\ref{sec:devmet}. The NW is modeled by a hexagon, with 40~nm length of edge (which corresponds to $W = 80$~nm). In electrostatic viewpoint a single conduction band is assumed with parabolic dispersion, and hard-wall confinement potential.  Fermi level pinning at the surface of the NW gives rise to band bending \cite{SmitJVSTB1989,Lueth1989}, which for simplicity is neglected in our model.
 Furthermore we fixed the Fermi energy at the edge of the conduction band for zero gate voltages.

In the simulation the electric potential, $V(\vec r)$ was calculated by solving the Poisson-equation,
\bnen \label{eq:poi} \Delta V(\vec r) = \frac{\rho(\vec r)}{\epsilon_0\epsilon^{i}_r}, \eden
where, $\epsilon^{i}_r$ is the relative dielectric constant of medium $i$ ($\epsilon^{\textrm{SiO$_2$}}_{r}=3.9$, $\epsilon^{\textrm{InAs}}_{r}=15.15$), $\epsilon_0$ is the vacuum permittivity, and $\rho(\vec r)=-e n(\vec r)$ is the electron charge density, which is only non-zero within the NW. The electron density of the NW, $n(\vec r)$ is given by the bulk density of states (DOS) of the InAs at zero temperature,
\bnen \label{eq:edens} n(\vec r) = \frac{3}{4\pi^2} \left( \frac{2m^*}{\hbar^2} \right)^{3/2} \left[eV(\vec r)\right]^{3/2}, \eden
where $e$ is the electron charge, $\hbar=h/2\pi$ is the Planck constant, $m^*=0.023 m_e$ is the effective mass of the electrons in the conduction band, and $m_e$ is the free electron mass. In the simulation Eq.~\ref{eq:poi}\&\ref{eq:edens} are solved self-consistently in an iterative manner. The details of the simulation can be found in the Appendix.

The calculated electric potential and the DOS at the Fermi energy for the two type of measurements are shown in Fig.~3a-b\&d-e. The black lines represent the NW/SiO$_2$/vacuum interfaces, the scale is shown in Fig.~3a. The SG\&BG electrodes are not shown on the graphs, the SG1 (SG2) is on the left (right) 70~nm away from the edge of the NW, the BG is 400~nm below the NW/SiO$_2$ interface.

Fig.~3a\&b correspond to the asymmetric SG measurement with gate voltage values of $V_{SG1} = 11$~V, $V_{SG2} = -7.7$~V and $V_{BG} = 15$~V. The electron density (which can be directly calculated from the DOS with Eq.~\ref{eq:DOS}\&\ref{eq:fermi}) is concentrated to the left side of the NW, as a result of the attractive force of the positively charged SG1 electrode. For this gate configuration the numerical calculation yield an average electric field of 20~mV/nm inside the NW.

Fig.~3d\&e correspond to the BG measurement with grounded SG electrodes with $V_{SG1} = V_{SG2} = 0$~V and $V_{BG} = 58$~V gate voltage values. The electron density is concentrated to the bottom part of the NW. The average electric field within the NW is about 25~mV/nm for this gate configuration.

The small areas of low DOS values at the surface of the NW in Fig.~3b\&e are artifacts arising from the grid of the simulation (5~nm).

The spin relaxation length associated to the external field-induced Rashba SOI can be calculated from magnitude of the electric field supplied by the simulation with the following formula \cite{BychkovRashba1984}:
\bnen \label{eq:lsor} l_{SO,R} = \frac{\hbar^2}{m^* e \alpha_0 \left\langle E \right\rangle}, \eden
where $\alpha_0 = 1.17$~nm$^2$ for InAs.  $\left\langle E \right\rangle$ is the spatially averaged electric field within the NW, weighted with the DOS at Fermi energy (see Eq.~\ref{eq:averE}). 

In the measurements the $l_{SO}$ was finite at zero gate voltages, which indicates that other sources of spin relaxation are present. To take them into account, we introduce the built-in spin relaxation length, $l_{SO,Bi}$. The built-in spin relaxation processes may originate from breaking the inversion symmetry of the system by the crystal structure, confinement potential, or scattering centers. For the sake of simplicity we assume that these processes are independent from the external gate voltages. For the different spin relaxation contributions the following sum rule was used:
\bnen l^{-2}_{SO} = l^{-2}_{SO,Bi} + l^{-2}_{SO,R} \eden
The value of $l_{SO,Bi}$ was chosen such that the measured and the calculated $l_{SO}$ values coincide with each other at zero asymmetric SG/BG voltage in the two demonstrated measurements. The simulation contains no further fitting parameters.

The measured and simulated $l_{SO}$ curves are shown in Fig.~3c\&f for the two different electrostatic field profiles. The measured $l_{SO}$ curves are the same as in Fig.~2a\&c. In both cases the simulated curves reproduce well the main findings of the measurement, that the SOI is enhanced by increasing electric field. Furthermore, despite the simplicity of the model and the low number of model parameters fairly good quantitative agreement was found.

In both cases $l_{SO}$ tends to saturate at higher voltages, which is the result of electrostatic screening by electrons. As it can be seen from the DOS graphs in Fig.~3b\&e, the electron density can reach high values in the NW close to the positively charged gate electrode. At higher electron densities the screening is more efficient, in agreement with the static screening theory, which leads to the decreasing enhancement of SOI as the gate voltages are further increased. To quantify the screening effect we estimate the screening length in the framework of Thomas-Fermi model \cite{AshcroftMermin}. In this model the screening length is given by
\bne r_{TF} = \left(\frac{e^2 D(\varepsilon_F)}{\epsilon_0 \epsilon_r}\right)^{-1/2}. \ede
With a typical value of DOS at Fermi level, $D(\varepsilon_F)=1\cdot 10^{31}$~eV$^{-1}$cm$^{-3}$, for InAs ($\epsilon^{\textrm{InAs}}_r=15.15$) the screening length is $r_{TF} \approx 10$~nm, which is consistent with the simulated electric potential profile (see Fig.~3a\&d).

We have shown that the introduced electrostatic model can explain our experimental results with an external field-induced Rashba SOI, assuming another, constant source of spin relaxation.

Following the standard framework of WAL \cite{HansenPRB2005} we assume that the SOI is the main mechanism of the spin relaxation and the measured spin relaxation length gives a good measure of the strength of the SOI. Using Eq.~\ref{eq:lsor} and the measured $l_{SO}$ values, one can estimate the strength of the SOI, $\alpha_R = \alpha_0 e \left\langle E \right\rangle = \hbar^2/m^* l_{SO}$, which is reasonable at high external fields, when the induced Rashba SOI dominates the built-in contribution. Tuning $l_{SO}$ from 150 to 70~nm corresponds to $\alpha_R \approx 2 \cdot 10^{-11}$~eVm and $4.3 \cdot 10^{-11}$~eVm, respectively.

In a very recent work of van Weperen et al. (see Ref.~\onlinecite{WeperenPRB2015}) a more realistic model of WAL theory is presented, which takes the 3D structure of the NW into account more precisely and allows us to extract spin-orbit induced spin relaxation length $l_{SO,R}$ and the Rashba spin-orbit coupling parameter $\alpha_R$ more accurately. However, the parameter range of the model does not cover the condition of our measurements since for our sample $W/l_e$ is around $4-8$. We believe that additional model calculation would be desired in order to give an accurate estimation of the SOI strength $\alpha_R$, which goes beyond some simplifications: As the source of spin relaxation mechanism most models consider a simple Rashba type SOI induced by a homogenous electrostatic field in the entire NW. In our device the electrostatic field is not homogeneous and in addition other SOI term, like D'yakonov-Perel has to be taken into account due to lack of inversion symmetry of wurtzite NW structure. Nevertheless it does not influence our finding that $l_{SO}$ can be tuned by gate voltages even if the conductance and the average density is kept constant.

	\begin{figure}[!htbp]
	\begin{center}
	\includegraphics[width=8.5cm]{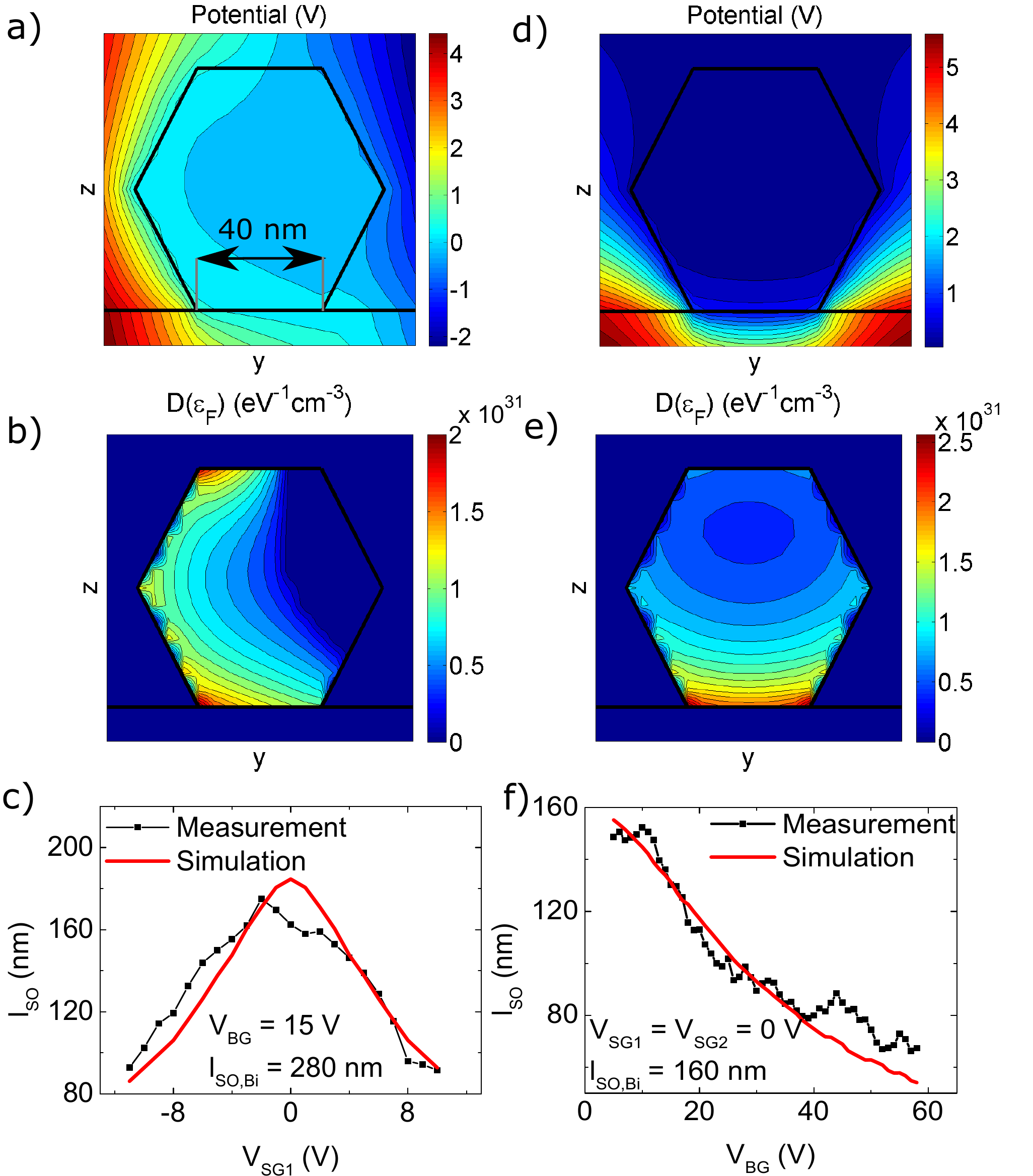}
	\caption{Results of the numerical simulation.
	a),b) The electric potential and density of states (DOS) at Fermi energy, respectively, in the vicinity of the nanowire (NW), for $V_{SG1} = 11$~V, $V_{SG2} = -7.7$~V, $V_{BG} = 15$~V. The black lines mark the NW/SiO$_x$/vacuum boundaries. The low DOS values at the surface of the NW is the result of the relatively coarse resolution of the grid (5~nm).
	c) Spin relaxation length in the asymmetric side gate configuration, the measurement data is same as in Fig.~2a (black squares), and the result of the numerical calculation (red curve).
	d),e) Same as a) and b), for BG-tuning measurement with $V_{SG1} = V_{SG2} = 0$~V, $V_{BG} = 58$~V.
	f) Same as c), for tuning with BG.}
	\label{fig3}
	\end{center}
	\end{figure}

\section{Conclusion}

The SOI of InAs NWs was studied by WAL measurements in a side gated geometry. We have shown that the strength of the SOI can be enhanced by up to a factor of 2, applying electric field across the NW. The tuning was demonstrated at a nearly constant conductance with asymmetric voltages applied on opposite SGs. Furthermore, a similarly large tuning was observed applying a high BG voltage, which strongly changed the electron density as well.

An electrostatic model was introduced to calculate strength of Rashba SOI induced by the external electric field. On the comparison of the experimental results and the numerical calculations good agreement was found, which supports that the Rashba SOI is the origin of gate-induced changes of the WAL signal. The possibility to tune the Rashba SOI strength in-situ without changing the electron density of the device is highly promising for various quantum electronic devices.

In order to gain a better insight into the SOI in NWs, a more realistic model calculation would be desired which takes into account the inhomogeneous electrostatic field in the NW, the presence of NW confinement potential, the band bending at the interface, the SOI term caused by the internal inversion asymmetry of the wurtzite crystal structure and the quantized conductance channels of the NWs. Such a more realistic theory could shine light on the limits of the validity of the simple 1D Rashba SOI Hamiltonian for the description of spin physics in semiconductor NWs.

\acknowledgments

We acknowledge useful discussions with Andras Palyi, Balazs Dora, Attila Geresdi, Peter Makk, Christian Schonenberger, Stefan Oberholzer, Samuel d'Hollosy, Balint Fulop and Thomas Sand Jespersen. We acknowledge support from EU ERC CooPairEnt 258789, FP7 SE2ND 271554,
Hungarian Grant No. OTKA K112918. S.C. was supported by the Bolyai Scholarship and G. F. was a SCIEX fellow (project NoCoNano)

\appendix

\section{Simulation}
\label{sec:sim2}

In this section the details of the electrostatic simulation are discussed.
%\red {Nem tul lenyeges: a 4. keplet utan van ertelme a 7 keplet ismetlesenek?}

The thickness of the SG electrodes is 100~nm, the separation of them is 220~nm. The thickness of the oxide layer is 400~nm. The boundary of the geometry is defined by a 1~$\mu$m~$\times$~1~$\mu$m square, which lower edge corresponds to the BG. The potential of the upper edge is fixed at zero potential. The potential of the right and left edges below (above) the SG electrode are defined by linear interpolation between the potential of BG and SG (SG and upper edge). Square lattice with 5~nm resolution is used in the simulations.

The NW is modeled with the properties of bulk InAs with 3 dimensional DOS
\bnen \label{eq:DOS} D(\varepsilon) = \begin{cases} \frac{1}{2\pi^2} \left( \frac{2m^*}{\hbar^2} \right)^{3/2} \sqrt{\varepsilon} & \text{if } \varepsilon \geq 0 \\ 0 & \text{if } \varepsilon < 0 \end{cases}, \eden
with hard-wall confinement potential, neglecting the band bending.

During the simulation the Poisson-equation (Eq.\ref{eq:poi}) and electron-filling,
\bnen \label{eq:fermi} n(\vec r) = \int^{\infty}_{0} D(\varepsilon)f(\varepsilon)d\varepsilon \eden
was calculated iteratively, where $f(\varepsilon)$, the Fermi-function was approximated with zero temperature form, the Heaviside step-function.

To assure the convergence of the algorithm for both the electron density and the potential the weighted average of the last two iteration step was used as new value, i.e.
\bean n^{(i)} &\rightarrow& (1-\eta_n) n^{(i-1)} + \eta_n n^{(i)} \nonumber \\ V^{(i)} &\rightarrow& (1-\eta_V) V^{(i-1)} + \eta_V V^{(i)}. \eean
In the simulations $\eta_n = \eta_V = 0.2$ was used.

To the calculation of SOI the electric field is calculated from the potential within the NW. Since only electrons at the Fermi-surface contribute to the conductance, the electric field is averaged with respect of the DOS at the Fermi-level,
\bnen \label{eq:averE} \left\langle E \right\rangle = \frac{\int d^2 r |\vec E(\vec r)| D(eV(\vec r))}{\int d^2 r D(eV(\vec r))} \eden

%\bibliographystyle{prsty}
%\bibliography{scheriff}

\end{document}